# Bulk Scheduling with the DIANA Scheduler

Ashiq Anjum, Richard McClatchey, Arshad Ali and Ian Willers, *Member, IEEE*

*Abstract*—Results from the research and development of a Data Intensive and Network Aware (DIANA) scheduling engine, to be used primarily for data intensive sciences such as physics analysis, are described. In Grid analyses, tasks can involve thousands of computing, data handling, and network resources. The central problem in the scheduling of these resources is the coordinated management of computation and data at multiple locations and not just data replication or movement. However, this can prove to be a rather costly operation and efficient sing can be a challenge if compute and data resources are mapped without considering network costs. We have implemented an adaptive algorithm within the so-called DIANA Scheduler which takes into account data location and size, network performance and computation capability in order to enable efficient global scheduling. DIANA is a performance-aware and economy-guided Meta Scheduler. It iteratively allocates each job to the site that is most likely to produce the best performance as well as optimizing the global queue for any remaining jobs. Therefore it is equally suitable whether a single job is being submitted or bulk scheduling is being performed. Results indicate that considerable performance improvements can be gained by adopting the DIANA scheduling approach.

## I. INTRODUCTION

In scientific environments such as High Energy Physics (HEP), hundreds of end-users may individually or collectively submit thousands of jobs that access subsets of the petabytes of HEP data distributed over the world and this type of job submission is known as bulk submission. Given the large number of jobs that can result from splitting the bulk submitted jobs and the amount of data being used by these jobs, it is possible to submit the job clusters to some scheduler as a unique entity, with subsequent optimization in the handling of the input datasets. In this process, known as bulk scheduling, jobs can compete for scarce compute and storage resources and this can distribute the load disproportionately among available Grid nodes.

Previous approaches have been based on so-called greedy algorithms where a job is submitted to a 'best' resource without assessing the global cost of this action. However, this can lead to a skewing in the distribution of resources and can result in large queues, reduced performance and throughput degradation for the remainder of the jobs. In contrast the familiar batch-system model for job execution is somewhat different in that the user is faced with long response times and a low level of influence which can be ineffective for bulk scheduling. Most of the existing schedulers normally deal individually with jobs, cannot handle the frequency of the (potentially millions of) jobs and cannot treat clusters of jobs as atomic units such as is required in bulk job scheduling. They also do not take into account network aware characteristics which are an important factor in the scheduling optimization of data intensive jobs. Contemporary schedulers cannot reorganize and scale according to evolving load conditions and in addition exporting and migrating jobs to least loaded resources is also non-trivial. In this paper we present for the first time a DIANA scheduling system which not only allocates best available resources to a job but also checks the global state of jobs and resources so that the strategic output of the Grid is maximized and no single user or job can undergo starvation. This scheduling system can efficiently exploit the distributed resources in that it is able to cope with the foreseen job submission frequency and is able to handle bulk job scheduling. In addition it takes into account network characteristics and data location and supports prioritization and multi-queuing mechanisms.

In this paper we introduce the DIANA Scheduling system and in particular its usage in scheduling bulk jobs. Section 2 introduces a case study and Section 3 describes related work in data intensive and network aware bulk scheduling. Section 4 explains the theoretical details of the scheduling decisions and Section 5 presents the scheduling algorithm. From section 6 onward we discuss the process for tackling bulk jobs. Section 7 illustrates the features of the bulk scheduling algorithm and section 8 the algorithm to handle bulk job scheduling. Section 9 describes the job migration algorithm to and Section 10 provides details of the queue management scheme. Finally

This work was supported in part by the Asia Link Programme of the European Commission under contract# ASI/B7-301/98/679/55(79286)

Ashiq Anjum is with the CCS Research Centre, University of the West of England, Coldharbour Lane, Bristol, UK BS16 1QY. (e-mail: 'ashiq.anjum@cern.ch').

Richard McClatchey is a Professor at the CCS Research Centre University of the West of England, Coldharbour Lane, Bristol, UK BS16 1QY (e-mail: richard.mcclatchey@cern.ch). (Richard McClatchey is the corresponding/submitting author.)

Arshad Ali is a Professor at the IT Institute of the National University of Sciences and Technology, Rawalpindi Pakistan. (e-mail: 'arshad.ali@niit.edu.pk').

Ian Willers is with the CMS computing Group at the CERN, European Organization for the Nuclear Research, Geneva Switzerland. (e-mail: 'ian.willers@cern.ch').

Section 11 describes our results. We show that a priority driven multi-queue feedback based approach is the most feasible strategy to facilitate bulk scheduling.

## II. CMS Data Analysis: A Case Study

We present a typical CMS physics analysis case to introduce the requirements, context and the problem domain that has been addressed in the DIANA system. CMS Physics analysis [1] is a collaborative process, in which versions of event feature extraction algorithms and event selection functions are iteratively refined until their physics effects are well understood. A typical physics job in an analysis effort might be `run this version of the system to identify Higgs events, and create a plot of particular parameters that have selected to determine the characteristics of this version'. The physicist normally runs the complete analysis in parallel by submitting hundreds or thousands of jobs accessing different data files. A job generally consists of many subjobs [2] and some large jobs might even contain tens of thousands of subjobs which can start and run in parallel. Each subjob consists of the running of a single CMS executable, with a run-time from seconds up to hours. The process may be multi-threaded, but in general the threads will only use the CPU power of a single CPU. Subjobs do not communicate with each other directly using an inter-process communication layer (such as MPI). Instead all data is passed, asynchronously, via datasets. Consequently if the data is concentrated on a single service, then this places a large burden on that service and the network to that service and this necessitates a special scheduling mechanism. A subjob generally has one or more datasets as its input, and will generally create or update at least one dataset to store its output. Within a job there is always an acyclic data flow arrangement between subjobs, regardless of how complex the subjob may be. This arrangement can be described as a data flow graph in which datasets and subjobs appear alternately. The data flow arrangement inside a job is known to the Grid components, in particular to the Grid Schedulers and execution services, so that they can correctly schedule and sequence subjob execution and data movement within a job.

Once the user has submitted the job to the Grid, the Grid Scheduler transforms the decomposed job description into a scheduled job description, which is then passed to the Grid-wide execution service. Often, the bulk of the CMS job output remains inside the Grid, as a new or updated dataset. However, one or more subjobs in a CMS Grid job might also deliver output (normally in the form of files) directly to the physics analysis tool that started the job; output delivery is asynchronous and should be supported by a Grid service. Presented below are the estimates [3] for the typical number of jobs from users and their computation and data related requirements which should be supported by the CMS Grid.

- Number of simultaneously active users: 100 (1000)
- Number of jobs submitted per day: 250 (10,000)
- Number of jobs being processed in parallel: 50 (1000)
- Job turnaround time: 30 seconds (for tiny jobs) - 1 month (for huge jobs) (0.2 seconds - 5 months)
- Number of datasets that serve as input to a subjob: 0-10 (0-50)
- Average number of datasets accessed by a job: 250,000 ($10^7$)
- Average size of the dataset accessed by a job: 30GB (1-3 TB)

Note that the parameters above have a wide range of values, so that simple averages are not very meaningful in the absence of variances. For each parameter, the first value given is the expected value that needs to be supported as a minimum by the Grid system to be useful to CMS. The second value, in parentheses, is the expected value that is needed to support maximum levels of usage by individual physicists. Given these statistics about workloads, it is clearly challenging to intelligently schedule tasks and to optimize resource usage over the Grid. This has led us to consider a bulk scheduling approach since simple eager or lazy scheduling models are not sufficient for tackling such distributed analysis scenarios.

## III. Related Work

Much work has been carried out in the domain of Grid scheduling however research in bulk scheduling for the Grid domain is relatively sparse. The European Data Grid (EDG) Project has created a resource broker under its workload management system based on an extended and derived version of Condor [4]. Although the problem of bulk scheduling has begun to be addressed (for example through the idea of shared sandboxes in the most recent versions of gLite from the EGEE project [5]), the approach taken is only one of priority and policy control rather than addressing real co-allocation and co-scheduling issues for the bulk jobs. In the adaptive scheduling scheme [6] for data intensive applications, Shi et al, calculate the data transfer cost for job scheduling. They consider a deadline based scheduling approach for data intensive applications and bulk scheduling is not covered. The Stork project [7] claims that data placement activities are equally important to computational jobs in the Grid so that data intensive jobs can be automatically queued, scheduled, monitored, managed, and even check-pointed as is done in the Condor project for computation jobs. Condor and Stork when combined handle both compute and data scheduling and cover a number of scheduling scenarios and policies however bulk scheduling functionality is not considered.

Thain et al. [8] describe a system that links jobs and data together by binding execution and storage sites into I/O communities. The communities then participate in the wide-area system and the Class Ad framework is used to express relationships between stake holders in communities; however again policy issues are not discussed. Their approach does cover co-allocation and co-scheduling problems but does not deal with bulk scheduling and how this can be managed through reservation, priority or policy. Basney et al. [9] define an execution framework linking CPU and data resources in the Grid in order to run applications on the CPUs which require access to specific datasets however they face similar problems in their approach to those discussed for Stork.

The Maui Cluster Scheduler [10] considers all the jobs on a machine as a single queue and schedules them based on a priority calculation. This approach assigns weights to the various objectives so that an overall value or priority can be associated with each potential scheduling decision, but it only deals with the compute jobs in a local environment. The data aware approach of the MyGrid [11] project schedules the jobs close to the data they require. However this traditional approach is not cost effective given the amount of available bandwidth in today's networks. The approach also results in long job queues and adds undesired load on the site when they could be moved to other least loaded sites. The GridWay Scheduler [12] provides dynamic scheduling and opportunistic migration but its information collection and propagation mechanism is not robust and in addition it has not as yet been exposed to bulk scheduling of jobs. The Gang scheduling [23] approach provides some sort of bulk scheduling by allocating similar tasks to a single location but it is tailored towards parallel applications working in a cluster whereas we are considering the Meta-Scheduling of the data intensive jobs submitted in bulk.

## IV. DIANA SCHEDULING

In this section we discuss the scheduling strategy of moving data to jobs (or both to a third location) and compare with the strategy of existing schedulers which always move the job to the data. One important drawback of existing schedulers is that network bottlenecks and execution or queuing delays can be produced in job scheduling. Data intensive applications often analyze large amounts of data which can be replicated over geographically distributed sites. If the data are not replicated to the site where the job is intended to be executed, the data will need to be fetched from remote sites. This data transfer from other sites can degrade the overall performance of job execution. If a computing job runs remotely, the output data produced needs to be transferred to the user for local analysis. To provide improvements in the overall job execution time and to maximize Grid throughput, we need to align and co-schedule the computation and the data (the input as well as the output) in such a way that we can reduce the overall computation and data transfer costs. We may even decide to send both the data and the executables to a third location depending on the capabilities and characteristics of the available resources.

We not only need to use the network characteristics while aligning data and computations, but we also need to optimize the task queues of the (Meta-)Scheduler on the basis of this correlation since network characteristics can play an important role in the matchmaking process and on Grid scheduling optimization. Thus, a more complex scheduling algorithm is required that should consider the job execution, data transfer and their correlation with various network parameters on multiple sites. There are three core elements of the scheduling problem which can influence scheduling decisions and which need to be tackled: data location, network capacity/quality and available computation cycles.

First we calculate the network cost. Network Losses are dependent on path conditions [13] and therefore the Network cost is:

Network Cost=Losses/Bandwidth

The second important cost which needs to be part of the scheduling algorithm is the computation cost. Paper [14] describes a mathematical formula to compute the processing time or compute cost of a job:

$$\frac{Q_i}{P_i} \times W5 + \frac{Q}{P_i} \times W6 + SiteLoad \times W7$$

Where $Q_i$ is the length of the waiting queue, $P_i$ is the computing capability of the site i and SiteLoad is the current load on that site. W5, W6 and W7 are weights which can be assigned depending upon the importance of the queue and the processing capability. The third most important cost aspect in data intensive scheduling is the data transfer cost:

Data Transfer Cost (DTC) = Input Data Transfer Cost + Output Data transfer cost + Executables transfer cost

Here we take three different costs for data transfer. The input data transfer cost is the most significant since most jobs take large amounts of input data which again depends on the network cost. Higher network cost will increase the data transfer cost and vice versa. Once we have calculated the cost of each stake holder, the total cost is simply a combination of these individual costs thus:

Total Cost = Network Cost + Computation Cost + DTC

The main optimization problem that we want to solve is to calculate the cost of data transfers betweens *sites* (DTC), to minimize the network traffic cost *between the sites* (NTC) and to minimize the computation cost of a job *within a site*. This total cost covers all aspects of the job scheduling and gives a single value for each associated cost, thus optimizing the Meta-scheduling decisions.

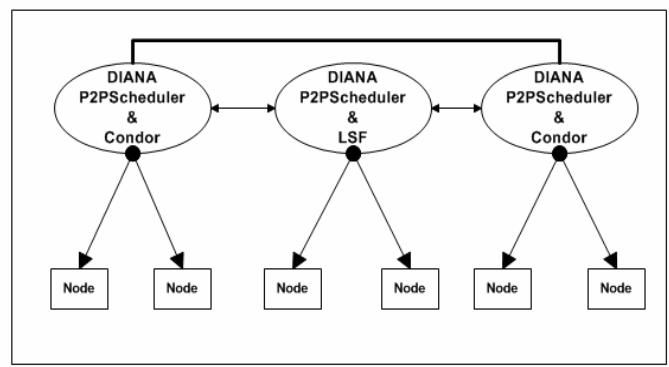

**Fig.1**: Communication between instances of Schedulers

In DIANA, we do not use independent Meta-Schedulers but instead use a set of Meta-Schedulers that work in a peer-to-peer (P2P) manner. As shown in Figure 1, each site has a Meta-Scheduler that can communicate with all other Meta-Schedulers on other sites. The Scheduler is able to discover other Schedulers with the help of a discovery mechanism [15]. We do not replace the local Schedulers; rather we have added a layer over each local Scheduler so that these local Schedulers can talk directly to each other instead of getting directions from a central global/Meta-Scheduler. In the DIANA architecture each local Scheduler has a local queue plus a global queue which is managed by the DIANA layer. This leads to a self organizing behaviour which was missing in the client server architecture.

## V. THE SCHEDULING ALGORITHM

This Scheduler deals with both computational jobs as well as data intensive jobs. In the DIANA Scheduling scheme, the Scheduler consults its peers, collects information about the peers including network, computation and data transfer costs and selects the site having minimum cost. To schedule computational jobs, this algorithm selects resources which provide most computational capability. The same is the case with data intensive jobs. To schedule data intensive jobs, we need to determine those resources where data can be transferred cost effectively. Since we have calculated the different costs, we can bring these costs under a scheduling algorithm as described below.

In the case of a computational job, more computational resources are required and the algorithm should schedule a job on the site where the computational cost is a minimum. At the same time, we have to transfer the job's files so we need to ensure that the job can be transferred as quickly as possible. Therefore, the Scheduler will select the site with minimum computational cost and minimum transfer cost. In the case of a data intensive job, our preferences will change. In this case our job has more data and less computation and we need to determine the site where data can be transferred more quickly and at the same time, where computational cost is also a minimum (or up to some acceptable level). The algorithm keeps on scheduling until all jobs are submitted. After every job we calculate the cost to submit the next job. The algorithm is as follows:

```
If the job is compute intensive then
   computationCost[] = getAllSitesComputationCost();
    NetworkCost []= getAllSitesNetworkCost();
   arrageSites[] = SortSites(computationCost, NetworkCost);   //it will sort array in ascending order
     for i=1 to arrangeSite.length
       site = arrangeSite[i]
        if ( site is Alive)
           send the job to this site
     end loop
end if
Else if the job is data intensive then
   dataTransferCost[] = getAllSitesDataTransferCost();
   NetworkCost []= getAllSitesNetworkCost();
   arrageSites[] = SortSites ( dataTransferCost, NetworkCost );   //it will sort array in ascending order
     for i=1 to arrangeSite.length
       site = arrangeSite[i]
        if ( site is Alive)
           send the job to this site
     end loop
end else-if
Else if ( job is dataintensive and compute intensive)
   computationCost[] = getAllSitesComputationCost()
   dataTransferCost[] = getAllSitesDataTransferCost()
  NetworkCost []= getAllSitesNetworkCost();
// since length of computationCost and dataTransferCost array is same. So we can use any of them
   siteTotalCost [] = new Array[computationCost.length]
   for i = 1 to computationCost.length
      siteTotalCost [i] = computationCost[i] + dataTransferCost[i] + NetworkCost [i]
   end loop
   sites [] = SortSites(siteTotalCost)
   for j = 1 to sites.length
     site = sites[i]
    if ( site is alive)
       schedule the job to this site
   end loop
```

## VI. PRIORITY AND BULK SCHEDULING

We describe here characteristics which can help us in creating an optimized scheduling algorithm. Clearly we want the jobs to be executed in the minimum possible time. One measure of work is the number of jobs completed per unit time i.e. the throughput. The interval from the time of submission to completion is termed the turnaround time and has significant bearing on performance indicators. Turnaround time is the sum of the periods spent waiting to access memory, waiting in the ready queue, executing the CPU and performing input/output. The waiting time is the sum of the periods spent waiting in the ready queue.

In an interactive system, turnaround time may not be the best criterion. Another measure is the time from the submission of a request until the first response has been provided. This measure, called the response time, is the time it takes to start responding but not the time that it takes to output that response. In the proposed DIANA algorithm, we aim to minimize the execution time, turnaround time, waiting and response time and to maximize the throughput.

### A. Priority based Scheduling

The proposed scheduling algorithm is termed a priority algorithm. A priority is associated with each process and the CPU is allocated to the process with the highest priority. Equal priority processes are scheduled on a First Come First Served (FCFS) basis. We discuss scheduling in terms of high priority and low priority. Priorities can be defined either internally or externally. Internally defined priorities use some measurable quantities to compute the priority of a process. For example, time limits, memory requirements, the number of open files and the ratio of I/O to CPU time can be used in computing priorities. External priorities are set by criteria that are external to the scheduling system such as the importance of the process. Priority scheduling can be either pre-emptive or non pre-emptive. The bulk scheduling algorithm described here is not a pre-emptive one; it simply places the new job at the head of the ready queue and does not abort the running job. Due to the interactive nature of most of the jobs, we follow a non pre-emptive mode of scheduling and execution. Since most jobs are data intensive, this makes it increasingly important to consider the non pre-emptive mode as a primary approach. A 'Round Robin' approach inside queues is not feasible in this case since most of the analysis jobs are interactive and the user is eagerly awaiting the output. Any delay in the output may lead to a dissatisfied user and we need to provide resources until the output can be seen. This approach also leads to the conclusion that the pre-emptive approach is not feasible for interactive jobs but can be considered for batch jobs. In this algorithm we consider only the interactive jobs used for a Grid-enabled analysis.

### B. MULTILEVEL Queue Scheduling

Due to the different quality of service requirements by the community of Scientific Analysis users, jobs can be classified into different groups. For example, a common division is made between interactive jobs and batch jobs. These two types of jobs have different response-time requirements, and so might have different scheduling needs. In addition, interactive jobs may have priority over batch jobs. A multilevel queue-scheduling algorithm partitions the ready queue into multiple separate queues.

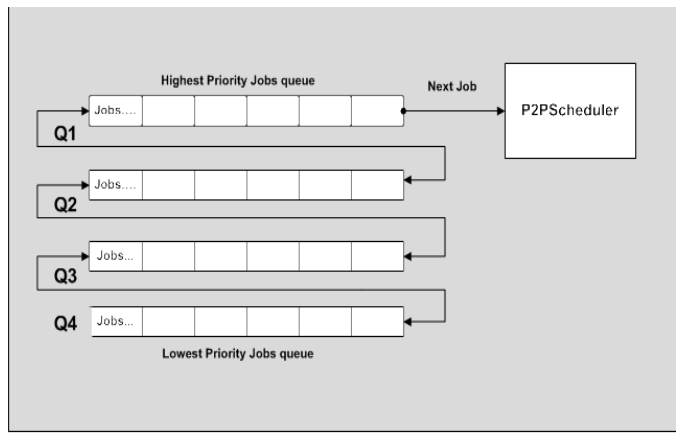

**Fig 2:** Multilevel feedback queues

In a multilevel queue-scheduling algorithm, jobs are permanently assigned to a queue on entry to the system. Jobs do not move between queues and this can create starvation if the jobs running are long duration jobs. We have employed multilevel feedback queue scheduling as shown in Figure 2 since it allows a job to move between queues. The idea is to separate processes with different requirements and priorities. If a job uses too much CPU time or is very data intensive, it will be moved to a higher-priority queue. Similarly, a job that waits too long in a lower-priority queue may be moved to a higher-priority queue.

### VII. BULK SCHEDULING ALGORITHM CHARACTERISTICS

We propose a multilevel feedback queue and priority-driven scheduling algorithm for bulk scheduling and its salient features are now briefly discussed. High priority jobs are executed first and the priority of jobs starts decreasing if the number of jobs from a user/site increases beyond a certain threshold. The priority becomes less than all the jobs in the queue if the job frequency is very high. A priority scheduling algorithm may leave some low priority processes waiting indefinitely for the CPU and we use an aging technique to overcome this starvation problem. Starvation of the resources is controlled by controlling the priority of the jobs. If no other job is available in the queue then all jobs from the user/site will be executed as high priority jobs. We do not employ quota and accounting since this restricts the users to a particular limit. Instead we use priority to schedule bulk jobs and to control the frequency as well as the queue on this basis. Similarly we do not follow the budget and deadline method of economy-based scheduling since the Grid is dynamic and volatile and the deadline method is feasible only for static types of environment.

All of the bulk jobs in a single burst will be submitted at a single site. If data and computing capacity is available at more than one site, we can consider job splitting and partitioning. Queue length, data location, load and network characteristics are key parameters for making scheduling decisions for a site. The priority of the burst or bulk of jobs is always the same since each batch of jobs has the same execution requirements.

Job migration between priority queues is a key point of the algorithm. Jobs can move between low priority to high priority queues depending upon the number of jobs from each user and the time passed in a particular low priority queue. Although migration of jobs between queues is supported within a single queue, we use the FCFS algorithm. Before jobs are placed inside the queue for execution, the algorithm arranges the jobs using the Shortest Job First (SJF) algorithm. We use the number of processors required as a criterion to decide between short or long execution times. Fewer processors required means job execution time is shorter and the job priority should be set higher. All shorter jobs are executed before longer jobs; this reduces the average execution time of jobs.

Priorities can be of three types: user, quota and system centric. We employ a system centric policy (embedded inside the Scheduler) since otherwise users can manipulate the scheduling process. In this manner a uniform approach will be set by the Scheduler for all users and a similar priority will be applied to all stake holders. Knowing the job arrival rates and execution capacity, we can compute utilization, average queue length, average wait time and so on. As an example, let $N$ be the average queue length (excluding the jobs being serviced), let $W$ be the average waiting time in the queue, and let $R$ be the average arrival rate for new jobs in the queue. Then, we expect that during the time $W$ that a job waits, $R*W$ new jobs will arrive in the queue. If the system is in a steady state, then the number of jobs leaving the queue must be equal to the number of jobs that arrive hence:

$$N = R*W$$

This equation, known as Little's Formula [16], is valid for any scheduling algorithm and arrival distribution. When a site is assigned too many jobs, it can try to send a number of them to other sites, which have more free resources or are processing fewer jobs. In this case, the jobs move from one site to another based on the criteria described in section IV. Once a job has been submitted on a remote site, the site at which it arrives will not attempt to schedule it again on another remote site (thus avoiding the situation in which a job cycles from one site to another). To each site we submit a number of jobs and a job reads an amount of data from a local database server, and then processes the data. If a site becomes loaded and jobs need to be scheduled on a remote site, the cost of their execution increases since the database server is no longer at the same site. If the amount of data to be transferred is too large or the speed of the network connections is too low, it might be better not to schedule jobs to remote sites but to schedule them for local execution.

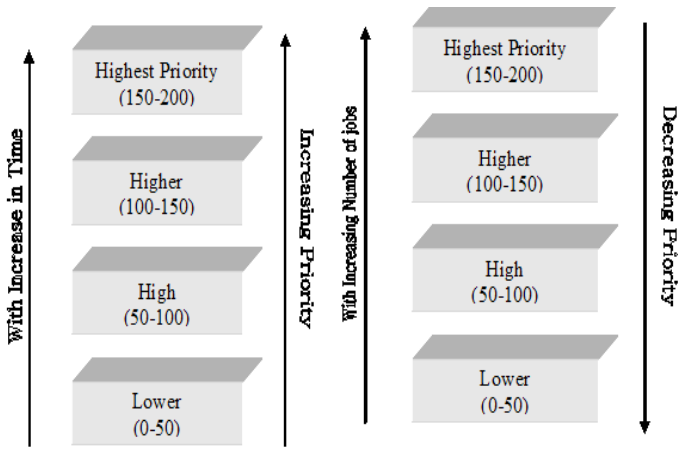

**Fig3:** Priority with Time and Job Frequency

In bulk scheduling there is a time threshold and a job threshold. If the number of jobs submitted from a particular user increases beyond the job threshold then the priority of the jobs submitted above the threshold number is decreased and jobs are migrated to a lower priority queue. In other words, with an increasing number of jobs, the priority of jobs from a particular user starts to decrease. Moreover, a time threshold is included to reduce the aging affect. With the passage of time, the priority of jobs in the lower priority queues is increased so that it can also have a chance of being executed after a certain wait time. In other words, the more time a job has to wait the more its priority continues to increase. This is illustrated in figure 3.

## VIII. BULK SCHEDULING ALGORITHM

We take each bulk submission of jobs from a user as a single group. Each group is taken as a single Job by the Meta-Scheduler which is scheduled by the DIANA algorithm of section IV. If this group is too large to be handled by a site, it is divided into subgroups, each having a sizeable number of jobs which can be handled by any number of the sites in the Virtual Organization (VO). The VO administrator sets the size of the subgroups which are created if the size of the group is very large and cannot be accommodated by any single site. This size varies from one VO to another. We assume that jobs are divided into equal but relatively smaller subgroups. The size of the subgroup is again set by the VO administrator. The size of the group is specified in the job description language file.

First the Scheduler checks whether the size of the group can be handled by a single site or not. Even if there is a site which can handle the whole group, it still checks whether it is cost effective to place this group on that particular site or whether it is more cost effective to divide the group into subgroups and submit the resulting subgroups to different sites. While placing the group or its subgroups, the DIANA scheduling algorithm is used and each group/subgroup is treated as a single job for the Meta-Scheduler. If the whole group is scheduled to a single site then the whole result is returned to the location which was specified by the user. In the case of subgroups, all the data from the subgroup execution sites is aggregated to a user specified location. No two groups from a single user or from different users can become part of a single group during the scheduling. Each group from each user maintains its identity and is treated independently by the Scheduler. The pseudo code of the algorithm is as follows:

Set the size of group filed in the jdl.
Set the group division factor
Submit the bulk Job in groups
Get list of sites
Check the queue size and computing capacity of each site
Check the data location and data requirements of the group
Match the site capacity against the bulk job group
Use the DIANA scheduling approach to select a site

If whole group can be accommodated by the site
    Submit the group to that site
    Aggregate the output of all jobs in the group
    return the results to the user's specified location
else
    Divide the group into subgroups using the group division factor
    Find the matching sites for the subgroups
    Submit each group to different site using DIANA scheduling technique
    Aggregate the out put of all the subgroups
    return the results to the user's specified location

For example, the user submits 10,000 jobs in a bulk job. Let us suppose, there are four sites A, B, C and D having 100, 200, 400 and 600 CPU's respectively. We assume that the network and data conditions of all three sites are the same. Since these are bulk jobs, they have similar characteristics and we assume that each job in the group takes one hour to get processed. Using the algorithm stated above, we can have three possibilities. Either to submit all the jobs on a single site, to divide the jobs into two best sites (in our case C and D) or to divide the jobs into four sites. The following table gives the times taken in each process.

| Jobs | Group | A (100) | B (200) | C (400) | D (600) | Total execution Time (hours) |
|---|---|---|---|---|---|---|
| 10,000 | 1 | | | | 10,000 | 16.6 |
| 10,000 | 2 | | | 4,000 | 6,000 | 10 |
| 10,000 | 10 | 1,000 | 2,000 | 3,000 | 4,000 | 8.5 |

**Fig4:** Job groups and execution improvements

From the table in Figure 4 we can see that by dividing the jobs into a number of groups, the Scheduler has clearly optimized the job executions times. Smaller job groups mean greater optimization. Moreover shorter jobs get higher priorities as discussed earlier and therefore there are greater chances of their earlier execution and this further optimizes the scheduling process. This also gives the advantage of including smaller sites into the execution process which otherwise will remain underutilized.

There can also be a job execution limit on a site so that a user cannot execute more than a fixed number of jobs. This concept of small groups will clearly also help to optimize the scheduling process. Furthermore there are certain large sites where, at a single point in time, all the processors might not be available and all the remaining available computing capability can be utilized by assembling small groups. This will reduce the queue as well as the load on the large sites and will also provide room for the high priority jobs to be executed. However this does not necessarily mean that just computing power is taken into account as a submission criterion. Each group of jobs is submitted using the DIANA scheduling algorithm which ensures that only that the site which has the least overall cost for its execution is selected for a group or a single job. We also described earlier that SJF execution reduces the average execution times of the all the jobs and this principle is also applicable here. In the case of larger groups, the waiting times for jobs will be longer and this will affect the overall execution time. Small groups will spend less time in the queue by getting higher priorities and therefore overall execution time will be further reduced.

## IX. JOB MIGRATION ALGORITHM

To illustrate job migration let us take an example scenario where a user submits a job to the Scheduler and the Scheduler puts this job into queue management. If the queue management algorithm (see section VII) of the Scheduler decides that this job should remain in the queue, it may have to wait a considerable time before it gets serviced or before it is migrated to some other site. In this case the queue management module will ask the scheduling module to migrate the job. The important point to note here is that we want the job to be scheduled at that site where it can be serviced earliest. Therefore our peer selection criteria is based on two things: the minimum queue length and the minimum cost to place this job on the remote site.

The Scheduler will communicate with its peers and ask about their current queue length and the number of jobs with priorities greater than the current job's priority. The site with minimum queue length and minimum total cost is considered as the best site to where the job can be migrated. The algorithm will work as follows

*Sites[] = GetPeerList( )*
*int count = Sites.length // total no of sites*
*int queueLength [ ] = Sites.length*
*int jobsAhead[]= new int[ count ]*
*for ( i=1 to count )*
*   jobsAhead [i] = getJobsAhead( site[i] )*
*end for*
*find the peer with minimum jobsAhead*
*if ( peer's jobsAhead < localsite's jobsAhead) then*
*increase the job's priority*
*migrate the job to that site*
*else*
*   keep the job on local site*

First it will get the information about the available peers from the discovery or information service. Then it will communicate with each peer and collect the peer's queue length, total cost, and the number of jobs 'ahead' of the current job's priority. After this, it will find out the site with the minimum queue length and minimum jobs ahead. If the number of jobs and total cost of the remote site is more than the local cost, then this job is scheduled to the local site. In this case the other sites are already congested and there is no need to migrate the job. Therefore that job will remain in the local queue and will be served when it gets the execution slot on the local site. Otherwise the job is moved to a remote site subject to the cost mechanism. This decision is made on the principle that this job as a result will get quicker execution since the targeted site has overall least cost and least queue as compared to other sites.

This policy is not just all-to-all communication. The nodes are divided into SubGrids, each SubGrid having its own "RootGrid". Roughly each site has one RootGrid and may have one or more SubGrids. The Meta-Scheduler works at the RootGrid (Master node) level in this approach and therefore we use the RootGrid, Master and Meta-scheduler interchangeably to describe this approach. The RootGrid to RootGrid communication is in essence a P2P communication between the Meta-schedulers. Each RootGrid maintains a table of entries about the status of the nodes which is updated in real time when a node joins or leaves the system. Local schedulers work at the SubGrid level. When a user submits a job, the Meta-Scheduler at the RootGrid communicates within the SubGrid to find suitable resources. If the required resources are not available within the SubGrid, it contacts the RootGrids of other SubGrids in the VO which have suitable resources. Therefore a single machine within a SubGrid communicates only with the Meta-scheduler, which itself communicates with the Meta-schedulers at other RootGrids. Consequently, this approach is not just all-to-all communication.

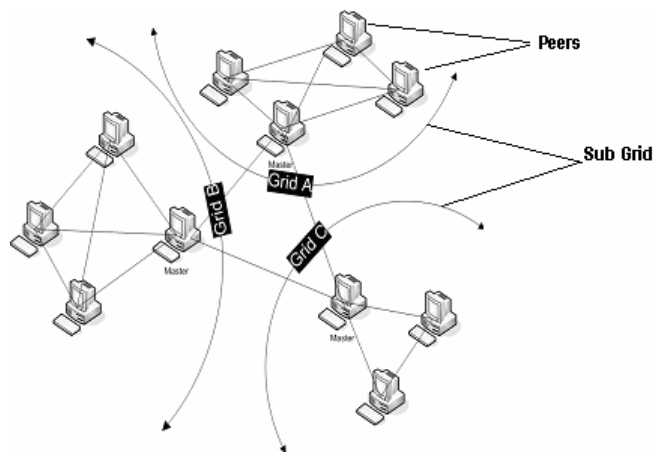

**Fig 5**: Topological Structure

A RootGrid contains all information about the nodes in its SubGrid. In case a RootGrid crashes, a standby node in the SubGrid can take over as a RootGrid. The RootGrid replicates its information to this standby node to avoid information loss. The RootGrid should always be the machine with the largest availability within that SubGrid and will have a unique ID, which will be assigned at the time of its joining the Grid. After joining, a Peer will check for the existence of the RootGrid. If the RootGrid does not exist, it means this is the first Peer joining the system. That Peer will then create the RootGrid and will join it. If the RootGrid exists then the Peer will automatically join that RootGrid and will search for its SubGrids and will join the nearest SubGrid using the criteria stated earlier. Whenever a site becomes part of the Grid, a separate SubGrid encompassing the site resources is created which joins the nearest RootGrid. If the site is fairly small in terms of the resources, this site may also join some existing SubGrid. The size of the SubGrid and RootGrid and other policy decisions have to be taken by a VO administrator and may vary from one Grid deployment to another. This algorithm will setup the topology, as shown in Figure 5.

## X. Queue Management

We propose a multi queue feedback-oriented Queue Management in which jobs are placed in the queues of varying priority. Each queue will contain jobs having priorities that fall in its specified priority range. According to our priority calculation algorithm, the priority of all the jobs will be in the interval {-1, 1} where -1 indicates the lowest priority and 1 indicates the highest priority. Therefore, the priority ranges for four proposed queues (Q1, Q2, Q3, and Q4) is proposed to be:

$Q1: 0.5 \leq priority < 1$
$Q2: 0 \leq priority < 0.5$
$Q3: -0.5 \leq priority < 0$
$Q4: -1 \leq priority < -0.5$

In the process of selecting a job's position in the queue, we place the jobs in the descending order of their priorities i.e. the job with the highest priority will be placed first in the queue and a priority order is followed for the rest of jobs. Finally we determine all those jobs having the same priority, and arrange them on a FCFS basis. Job migration between queues is an essential feature of our Queue Management. On the arrival of each new job, all the jobs already present in the queues are re-prioritized. The re-prioritization algorithm may result in the migration of jobs from low priority to high priority queues or from high priority to low priority queues. The re-prioritization technique militates against aging since jobs are assigned new priorities on the arrival of each new job and each job gets its appropriate place in the queues according to the new circumstances. In the case of congestion in the queues, the Queue Management algorithm will migrate the jobs to any other remote site where there are fewer jobs waiting in the queues. However, only low priority jobs are migrated to remote sites because low priority jobs (e.g. for a job falling in Q4) will have to wait for a long time in the case of congestion.

Knowing the arrival rate and the service rate of the jobs, we can decide whether to migrate the job to some other site or not. The formula to decide whether there is congestion in the queues or not is:

*If(Arrival Rate – Service Rate ) / Arrival Rate > Thrs*

where Thrs is the threshold value configurable by the administrator. If we increase Thrs, then this has the effect that the arrival rate may exceed the service rate and we must allow more jobs in the queues and consequently there is less migration. In any case this value lies in the {0, 1} interval. Taking this, we can now explain the queue management algorithm.

Suppose **'n'** is the total number of jobs of the user in all job queues, including any new job. Let the new job require 't' processors for the computation and let **'T'** be the total number of processors required by all the jobs present in all job queues. We denote the quota of the user, submitting the new job, by **'q'** and the sum of the quotas of all the users, currently having their jobs in the job queues including '**q**', by '**Q**'. So if the new user has already some jobs in the job queues, '**q**' will appear just once in the '**Q**'. Let **'L'** be the sum of lengths of all job queues i.e. the total number of jobs present in all job queues including the new job. Therefore if there are already, say, 15 jobs in the job queues when a new job arrives, then L would be 16. To assign a new job a place in the job queue, we associate a number to it. This number is called the "Priority" of the job and has its value in the interval {-1, 1}. The rule is that "the larger the priority, the better the place will be". Obviously if its priority is in the range {0,1}, it will be considered as favoured for execution. To attain a good priority we must meet the following two constraints:

$$\frac{n}{L} \leq \frac{q}{Q} \text{ and } \frac{1}{L} \leq \frac{t}{T} \quad \ldots\ldots\ldots (IV)$$

Or $n \leq \frac{(q \times L)}{Q}$ and $L \leq \frac{T}{t}$ ..........(V)

Combining these two inequalities IV and V, we get

$$n \leq \frac{(q \times T)}{(Q \times t)} \quad \ldots\ldots\ldots (VI)$$

We denote $\frac{(q \times T)}{(Q \times t)}$ by '**N**'

'**N**' represents the threshold and obviously, it is dynamic. For each job, its value will be different. If a user's number of jobs in the queue crosses this threshold then the priority of the jobs crossing the threshold '**N**' must be lowered. To calculate the priority of the new job, we use the following algorithm:

*If ( n <= N )*
 *Pr(n) = (N – n) / N*
*Else*
 *Pr(n) = (N – n) / n*

where Pr (n) denotes the priority of the new job. Note also that the priority will always lie in the interval {-1, 1}.

On the arrival of each job, the priorities of all the other jobs will be recalculated. This technique is known as Reprioritization. The reason for doing this is that we want to make sure that the jobs encounter minimum average wait time and the most 'deserving' job in terms of quota and time is given the highest priority. Moreover, by using this strategy we need not worry about the starvation problem and there is no aging since jobs are reprioritized on the arrival of each new job. The algorithm to reprioritize the jobs is the same as that mentioned above. The value of **q** for a particular user's jobs remains the same, **Q** and **T** remain the same for all the jobs, however, **t** is job specific and it may vary with each job. Therefore, the value of '**N**' differs for each job. By using the above mentioned formula, we can calculate the priority for all the jobs and place them in their respective queues.

Of course, if more than one job shares the same priority then the timestamp associated with each job is compared and the older job, which has spent more time in the queue, is placed before the new job. Also note that when a job is taken out for service the rest of the jobs need not be reprioritized.

| User | Quota (q) | Job's Processors Requirements (t) | Total # of Processors Required by all queues (T) | User's Total# of Jobs (n) | Total # of Jobs in all queues | Quota Sum off all distinct users (L) | Priority Pr(n) (Q) |
|---|---|---|---|---|---|---|---|
| A | 1900 | 1 | 7 | 2 | 3 | 3600 | 0.4586 |
| A | 1900 | 5 | 7 | 2 | 3 | 3600 | -0.6305 |
| B | 1700 | 1 | 7 | 1 | 3 | 3600 | 0.6974 |

**Fig6:** Priority calculation for jobs from different users

Let us consider an example scenario where a new job is submitted by user A and it requires one processor i.e., t = 1. We assume that the quota q for user A is 1900 and currently there is no job in the queue therefore, L =1, n = 1, Q = 1900 and T = 1 and N = (1900 * 2) / (1900 * 2). If we put these values in the algorithm and the test 'if' condition is true, then this job is placed in Q2.This scenario is shown in figure 6.

We assume that the first job has not as yet been serviced and meanwhile, user A submits his second job demanding five processors i.e. t = 5, then L=2, n = 2, T = 1 + 5 = 6, q = 1900, Q = 1900 and N = (1900 * 5) / (1900 * 3). Again putting these values in the algorithm, we find that the 'if' condition becomes false and Pr (n) = -0.4 and therefore the job is placed in Q3. Reprioritization then starts and the priority of the job already present in the queue is recalculated. This time the priority is set to 0.666666 and this job is migrated from Q2 to Q1 i.e., the highest priority queue as shown in the figure 6. This is of interest because user A has submitted only two jobs and the threshold has not been exceeded on the second job. The algorithm equally handles all users and jobs and the priorities decrease as the number of jobs by a user increases (and it does not matter that the second job exceeds the threshold). Now suppose that another user B submits his first job which requires one processor i.e. t=1 having user quota of 1700, q=1700. Assuming that the two jobs by user A are still in the queues, L=3, n=1, T= 1+5+1= 7, and Q = 1900+1700 = 3600. The 'if' condition holds true and Pr (n) = 0.6974 and therefore the job is placed in Q1. Reprioritization starts and as the result, the priorities of the previous jobs change and the first job by user A is migrated from Q1 to Q2 and the second job by user A is migrated from Q3 to Q4. This is illustrated in figure 6. It is notable that the first job by both user A and B demands one processor and the quota of user A is greater than user B, even if the priority of user B's job is greater than the user A job. This is because user A has submitted more jobs than user B and the algorithm handles this while calculating priorities. In this way the algorithm manages and updates the queues on the arrival of each new job.

## XI. RESULTS AND DISCUSSION

We present here results from a set of tests which have been conducted with the DIANA Scheduler, using a prototype implementation and MONARC [17] simulations to check the algorithm behaviour for bulk scheduling. We compare our experimental results with the EGEE work load management system. For simplicity we have used our own test Grid (rather than a production environment) to obtain results since a production environment requires the installation of many other Grid components that are superfluous for the tests. We have used five sites for the purpose of this experiment. Site 1 has four nodes and the remaining four sites have five nodes each. First we submitted a number of jobs which exceeded the processing capacity of the site and observed large queues of jobs which cannot be processed in an optimal manner. The bulk scheduling algorithm discussed above was used to migrate the jobs to other sites. The results suggest that as the number of jobs increases beyond the threshold limit, more and more jobs are migrated to other less loaded sites over time since the site selection is no longer optimal. In selecting a single site, we use DIANA so that all the network, compute and data related details are brought under consideration before the job placement on the selected site.

DIANA makes use of a P2P network to track the available resources on the Grid. The current implementation makes use of three software components for resource discovery: Clarens [18] as a resource provider/consumer, MonALISA [19] as a decentralized resource registry, and a peer-to-peer Jini network provided by MonALISA as the information propagation system. The DIANA instances can register with any of the MonALISA peers through the discovery service and different instances can directly interact with each other. We have employed PingER [20] to obtain the required network performance information since it provides detailed historical information about the status of the networks. It is a mature tool that integrates a number of other network performance measurement utilities to provide one-stop information for most of the parameters. It does not provide a P2P architecture but information can be published to a MonALISA repository to propagate and access it in a decentralized manner.

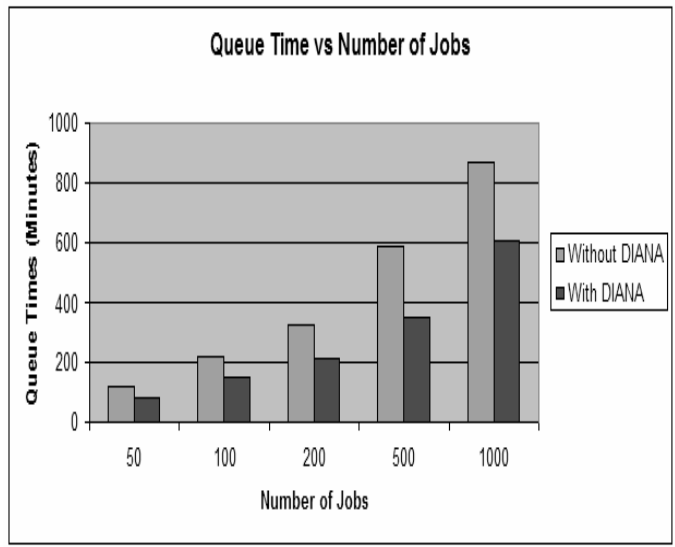

**Fig7**: Queue time versus number of jobs

The graphs in figures 7 show the optimization achieved by employing the DIANA algorithm. We can see that with an increasing number of jobs the execution performance increases. Here we note that DIANA is significant since as the number of jobs increases it finds only those sites for the job execution which are least loaded, which preferably have the required data and which have adequate network capacity to transfer the output data towards the client location. It is equally applicable to compute intensive jobs since it will find a site where having the shortest queue so that when the job is then placed it will get a higher execution priority than at its current execution site. Moreover the output data of the compute operation will be quickly transferred to the submission site due to the optimal selection of the link between the submission and execution nodes.

In tests we firstly submitted 25 jobs and observed their queue time and execution time. Then we submitted the same job three times and measured the queue and execution times once again. After this, we increased the number of jobs to 50 and then gradually to 1000, in order to check the capability of the existing matchmaking and scheduling system. The number of jobs was increased for two reasons. Firstly, to check how the queue size increases and secondly to determine in which proportion the Meta-Scheduler submits the jobs (i.e. whether jobs are submitted to some specific site or on a number of CPUs at different locations depending on the queue size and the computing capability).

We calculated and plotted the queue time and how it increases and decreases with the number of jobs. We observed that both queue and execution time have similar trends; this is due to the fact that DIANA selected those sites which can most optimally execute the jobs and where jobs do not have to wait for long times in the queue to be executed. The queue time is almost proportional to execution time since if the job is running and taking more time on the processor, the waiting time of the new job will also increase accordingly.

The queue time of local resource management systems is very significant in the Grid environment and takes a certain proportion of the job's overall time (see figure 7). Sometimes this is even greater than the execution time if the resources are scarce compared to the job frequency. We took only a single job queue in the Scheduler and we assumed that all jobs have the same priority. In fact, the job allocation algorithm being employed is based on a First Come First Served (FCFS) principle. The FCFS queue is the simplest and incurs almost no system overhead. The queue time here is the sum of the time in the Meta Scheduler queue and the time spent in the queue of the local resource manager.

The graph of the queue times when the number of the jobs changes is shown in Figure 7. It shows that the queue grows with an increasing number of jobs and that the number of jobs waiting for the allocation of the processors for execution also increases. The graph shown in Figure 8 is based on average values of time for varying number of jobs as mentioned earlier. Improvements in the queue times of the jobs due to DIANA Scheduling are also depicted in the same figure.

Similarly, we monitored the execution times of the jobs. The execution time is the wall clock time taken for a job that is placed on the execution node. It does not include queue time or waiting time. By increasing the number of the jobs, it is evident from Figure 8 that the average time to execute a job is increased. More competing jobs clearly mean more time for a specific job to complete.

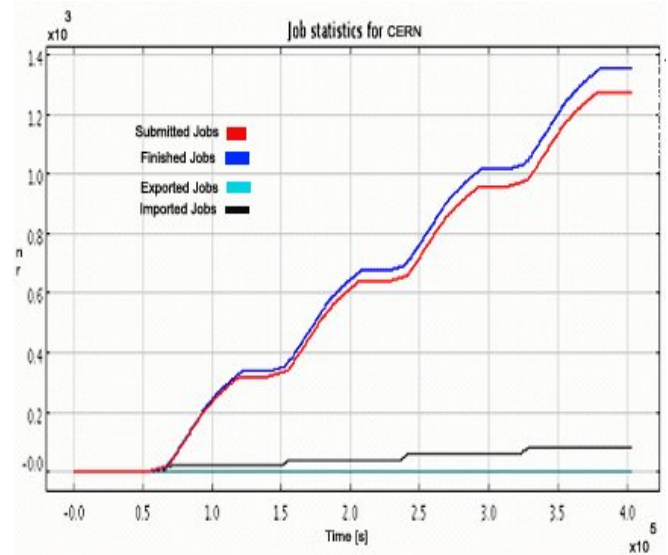

**Fig8:** Execution time versus number of jobs

Once jobs at a site exceed the threshold limit, the Bulk scheduling algorithm again uses the DIANA to select the best alternative site for execution in terms of computation power, data location, network capacity and queue length. As the number of jobs increase beyond a threshold, bulk scheduling algorithm employs policies and priorities to provide the desired quality of service to all or some preferred users and also restricts certain users making monopolistic decisions to avoid starvation for certain users.

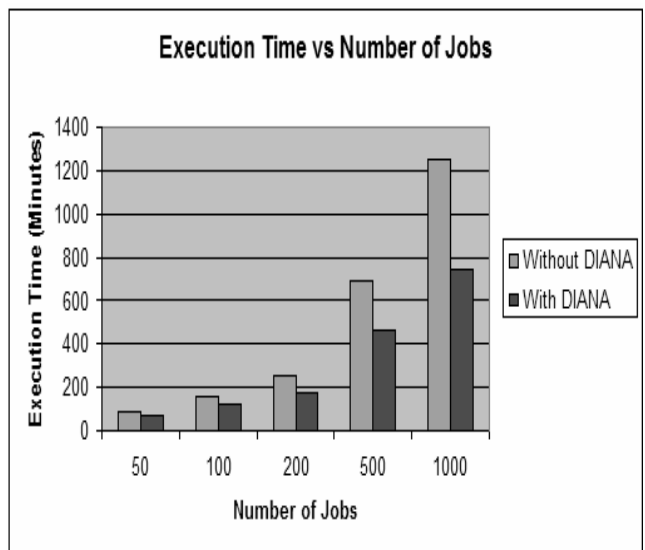

**Fig9**   Jobs execution and migration with Time

In figure 9 we can see the effect of jobs exceeding the execution capacity of a site and that jobs are exported to least loaded sites to optimize the execution process. Even the fluctuation in the submission rate is reflected by the corresponding export and execution rates. If the number of jobs being processed at a site is less than its execution capacity, then this site can import jobs from other sites in order to reduce the overall execution and queue time of jobs as shown in figure 10.

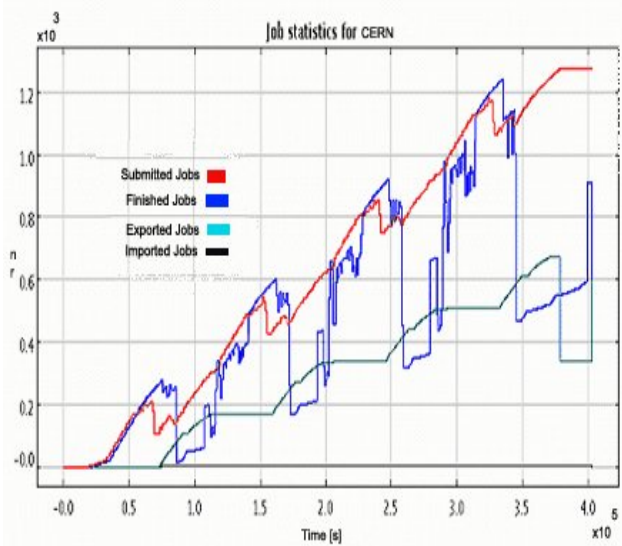

**Fig10:** Capacity of the site greater than the submitted jobs

If the Job submission frequency is much higher than the site consumption rate, the site keeps on processing the jobs at a constant rate and the rest of the jobs are exported to optimally selected sites. It is even possible for a site to export the jobs which do not have the required data locally as well as importing other jobs at the same time which can perform well locally and this is illustrated in figure 11. Figure 11 illustrates that the site is constantly executing the jobs at its peak capacity but at the same time the scheduler is migrating jobs which cannot perform well on this site to other optimal sites. Moreover, at the same it is also allowing the import of jobs from other sites which either have the required data available on this site or can get better execution priority or there is a shorter queue on this site compared to other sites. We are employing the non-pre-emptive approach in our bulk scheduling algorithm and once a job starts execution we do not move it since check-pointing [21] and re-start are very expensive operations in data intensive applications.

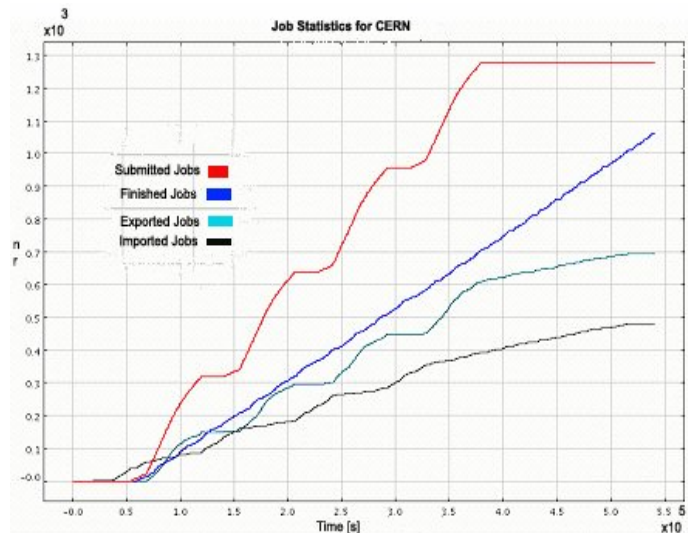

**Fig11** Job Frequency higher than the execution

## XII. CONCLUSIONS

In this paper we have studied the role of job scheduling in a data intensive and network aware Grid analysis environment and have proposed a strategy for job scheduling, queuing and migration. Our results indicate that a considerable optimization can be achieved using bulk scheduling and the DIANA scheduling algorithms for applications that are data intensive, such as those in large scale physics analysis. We presented here a theoretical as well as a mathematical description of the DIANA Meta-scheduling algorithms and it was shown that a scheduling cost based approach can significantly optimize the scheduling process if each job is submitted and executed after taking into consideration certain associated costs. This paper demonstrated the bulk scheduling capability of the DIANA Scheduler for data intensive jobs; further details can be found in [22] in which the cost based approach for scheduling is detailed but it does not cover the bulk scheduling process.

Queue time and site load, processing time, data transfer time, executable transfer time and results transfer time are the key elements which need to be optimized for optimization of scheduling and these elements were represented in the DIANA scheduling algorithm. The three key variables which need to be calculated were identified as data transfer cost, compute cost and network cost and were expressed in the form of mathematical equations. The same algorithm was extended and it was later demonstrated that if queue, priority and job migration were included in the DIANA scheduling algorithm, the same algorithm could be used for scheduling of bulk jobs. As a result, a multi-queue, priority-driven feedback based bulk scheduling algorithm is proposed and results suggest that it can significantly improve and optimize the Grid scheduling and execution process. This not only reduces the overall execution and queue times of the jobs but also helps avoid resource starvation.

Our approach is equally applicable to compute and data intensive jobs since compute intensive jobs, for example CMS simulation operations, also produce a large amount of data which needs to be transferred to the client location. Moreover, priority and queue management can significantly reduce the wait time of the jobs which in most cases is higher than the execution times. Similarly the data transfer time of jobs is reduced due to improved selection of the dataset replica while scheduling the job and this is further ensured by carefully evaluating the WAN link between the submission and the execution nodes. In conclusion this has helped to optimize the overall execution and scheduling process when either a single job is being executed or the bulk scheduling of jobs is being performed and this approach is equally applicable whether the jobs are compute or data intensive. The outcome of this work is being assessed for use in physics analysis chain of the Compact Muon Solenoid (CMS) project at CERN.

## XIII. References

[1] Data link Story: CMS Data Analysis Using Alliance Grid Resources, National Centre for Supercomputing Applications (NCSA) report, 2001.
[2] K. Holtman, on behalf of the CMS collaboration, CMS Data Grid System Overview and Requirements CMS note 2001.
[3] K. Holtman. HEPGRID2001: A Model of a Virtual Data Grid Application. Proc. of HPCN Europe 2001, Amsterdam, p. 711-720, Springer LNCS 2110. Also CMS Conference Report 2001/006.
[4] J. Frey et al., "Condor-G: A Computation Management Agent for Multi-Institutional Grids", Proceedings of the Tenth IEEE Symposium on High Performance Distributed Computing (HPDC10) San Francisco, California, August 7-9, 2001.
[5] P. Andreetto, S. Borgia, A. Dorigo, A. Gianelle, M. Mordacchini et.al, Practical Approaches to Grid Workload & Resource Management in the EGEE Project, CHEP04, Interlaken, Switzerland.
[6] X. Shi et al., "An Adaptive Meta-Scheduler for Data-Intensive Applications", International Journal of Grid and Utility Computing 2005 - Vol. 1, No.1 pp. 32 - 37
[7] T. Kosar and M. Livny, "A Framework for Reliable and Efficient Data Placement in Distributed Computing Systems", Journal of Parallel and Distributed Computing, 2005 - Vol 65 No. 10 pp. 1146-1157.
[8] D. Thain et al., M. (2001) "Gathering at the well: creating communities for Grid I/O", Proceedings of Supercomputing 2001, November, Denver, Colorado.
[9] J. Basney, M. Livny, and P. Mazzanti, "Utilizing widely distributed computational resources efficiently with execution domains", Computer Physics Communications, Vol 140, 2001.
[10] Brett Bode et al. The Portable Batch Scheduler and the Maui Scheduler on Linux Clusters 4th Annual Linux Showcase and Conference Atlanta, Georgia, October 2000.
[11] W. Cirne et al, "Running bag-of-tasks applications on computational Grids: The myGrid approach," in Proceedings of the ICCP'2003 – International Conference on Parallel Processing, October 2003.
[12] Eduardo Hudo, Ruben S. Montero, and Ignacio M. Llorente, "The GridWay Framework for Adaptive Scheduling and Execution on Grids " Scalable computing: practice and experience (SCPE) Volume 6, No. 3, September 2005.
[13] M. Mathis, J. Semke, J. Mahdavi & T. Ott, "The macroscopic behaviour of the TCP congestion avoidance algorithm", Computer Communication Review, 27(3), July 1997.
[14] H. Jin, X. Shi et al., "An adaptive Meta-Scheduler for data-intensive applications", International Journal of Grid and Utility Computing 2005 - Vol. 1, No.1 pp. 32 - 37
[15] A. Ali, A. Anjum, R. McClatchey, F. Khan, M. Thomas, A Multi Interface Grid Discovery System, Grid 2006, Barcelona Spain.
[16] J. H. Dshalalow, "On applications of Little's formula", Journal of Applied Mathematics and Stochastic Analysis, Volume 6, Issue 3, Pages 271-275, 1993.
[17] I. Legrand, H. Newman et al., "The MONARC Simulation Framework", Workshop on Advanced Computing and Analysis Techniques in Physics Research, Japan 2003
[18] C. Steenberg et al., ,The Clarens Grid-enabled Web Services Framework: Services and Implementation, CHEP 2004 Interlaken Switzerland and M. Thomas, et al., JClarens: A Java Framework for Developing and Deploying Web Services for Grid Computing ICWS 2005 , Florida USA, 2005.
[19] I. Legrand, MonaLIsa - Monitoring Agents using a Large Integrated Service Architecture, International Workshop on Advanced Computing and Analysis Techniques in Physics Research, Tsukuba, Japan, December 2003.
[20] L. Cottrell, W. Matthews. Measuring the Digital Divide with PingER, Second round table on Developing countries access to scientific knowledge, Trieste, Italy, Oct. 2003.
[21] K. Li, J. F. Naughton and J. S. Plank "Low-Latency, Concurrent Checkpointing for Parallel Programs", IEEE Transactions on Parallel and Distributed Systems, 5(8), August, 1994, pp. 874-879.
[22] A. Anjum, H. Stockinger, R. McClatchey, A. Ali, I. Willers, M. Thomas and F. van Lingen, "Data Intensive and Network Aware (DIANA) Grid Scheduling" Under final review at the Journal of Grid Computing, Springer Publishers 2006.
[23] P. Strazdins, J. Uhlmann, A Comparison of Local and Gang Scheduling on a Beowulf Cluster, cluster2004 San Diego, California